# 87. Star-Planet Interactions in the Radio Domain: Prospect for Their Detection

Philippe Zarka[1, 2]

(1) LESIA, Observatoire de Paris, CNRS, PSL, UPMC/SU, UPD, Meudon, France

(2) Observatoire Radioastronomique de Nançay, Observatoire de Paris, CNRS, PSL, Univ. Orléans, Nançay, France

**Philippe Zarka**
**Email:** philippe.zarka@obspm.fr

## Abstract

All types of interaction of a magnetized plasma flow with an obstacle (magnetized or not) are considered, and those susceptible to produce a radio signature are identified. The role of the sub-Alfvénic or super-Alfvénic character of the flow is discussed. Known examples in the solar system are given, as well as extrapolations to star-planet plasma interactions. The dissipated power and the fraction that goes into radio waves are evaluated in the frame of the radio-magnetic scaling law, the theoretical bases and validity of which are discussed in the light of recent works. Then it is shown how radio signatures can be interpreted in the frame of the cyclotron-maser theory (developed for explaining the generation of solar system planetary auroral and satellite-induced radio emissions) for deducing many physical parameters of the system studied, including the planetary or stellar magnetic field. Recent detections of such radio signatures with new generation low-frequency radiotelescopes and future prospects are then outlined.

## Introduction

Stars interact in many ways with their planets in orbit: through gravitation that constrains the orbit and, at short range – i.e., for close in exoplanets – causes tidal effects of each body upon the other (Cuntz et al. *2000*); through stellar light – especially at short wavelengths – that ionizes the planet's upper atmosphere (e.g., Encrenaz et al. *2004*); and through plasma and magnetic fields. This chapter is mainly interested with the latter, that may lead to the generation of intense radio emissions. The generic name "plasma interaction" is used for interactions involving plasma flows and magnetic fields.

# Plasma Flow-Obstacle Interactions

Various types of plasma interactions are observed in our solar system, involving the solar wind and magnetized or unmagnetized planets, as well as magnetized planets and their satellites. A coherent frame for sorting these various interactions is the general frame of flow-obstacle interactions (Zarka *2007*). The flow of plasma can be strongly or weakly magnetized (or unmagnetized in the limiting case), and sub-Alfvénic or super-Alfvénic. The obstacle is a conductive or insulating body, with or without an atmosphere, and possessing or not an intrinsic magnetic field (Lepping *1986*).

## Magnetized Obstacle

When the obstacle is magnetized, its interaction with the magnetized flow is likely to occur through magnetic reconnection at the flow-obstacle interface, where both magnetic field amplitudes are comparable but their orientations different (Dungey *1961*), leading to energy release in the form of plasma waves and particle acceleration susceptible to produce radio waves. The location in the system where radio emission may occur depends on the regions of space accessible to the accelerated electrons along magnetic field lines. In the case of a magnetized obstacle, these regions include the vicinity of the obstacle, usually toward the magnetic poles where its magnetic field lines converge, i.e. the auroral regions of planetary magnetospheres (magnetospheres are large-scale cavities carved in the solar wind by planetary magnetic fields (Bagenal *2001*; Encrenaz et al. *2004*)). There, accelerated electrons can collectively acquire non-Maxwellian distributions susceptible to produce intense radio waves. Those should also result from the interaction of stellar winds with magnetized exoplanets. The fact that the flow is strongly or weakly magnetized does not change the nature of the above phenomena. In the limiting and unlikely case where the flow is completely unmagnetized, it might still drive magnetospheric convection via viscous interaction (Axford and Hines, *1974*) and eventually electron acceleration and auroral radio emissions.

The perturbation caused by the obstacle to the magnetized plasma flow travels in the flow as Alfvén waves (carrying magnetic perturbations and associated electric currents along magnetic field lines), at the local Alfvén speed $V_A=B/(\rho\mu_o)^{1/2}$ with B the magnetic field amplitude, $\rho$ the local plasma density and $\mu_o$ the permeability of free space. The Alfvénic Mach number ($M_A = V_{flow}/V_A$) of the flow will define the topology and consequences of the interaction.

When the flow is super-Alfvénic ($M_A > 1$) in the obstacle's frame, a bow shock forms upstream of the obstacle that slows down and heats the flow, making it become sub-Alfvénic ($M_A < 1$) before it hits the obstacle and flows around it. Magnetic pile-up may occur on the obstacle. The shock "disconnects" the obstacle from the source of the flow and thus prevents Alfvén waves and thus most of the energy to travel back to this source. Some accelerated electrons at the shock or reconnection site may travel upstream but no corresponding prominent radio signature has been observed in the solar system.

When the flow is sub-Alfvénic, no shock is formed upstream of the obstacle and the Alfvén waves excited at the flow-obstacle interface can propagate along magnetic field lines toward the source of the flow. These Alfvén waves propagate in two "wings" in the plane that contains the flow velocity $\mathbf{V}_{flow}$ and magnetic field $\mathbf{B}_{flow}$ (if $\mathbf{V}_{flow}$ and $\mathbf{B}_{flow}$ are not parallel), oriented along the vectors ($\mathbf{V}_{A+} + \mathbf{V}_{flow}$) and ($\mathbf{V}_{A-} + \mathbf{V}_{flow}$), $\mathbf{V}_{A+}$ and $\mathbf{V}_{A-}$ being aligned along the direction of $\mathbf{B}_{flow}$ on both sides of the obstacle (Neubauer *1980*). Currents are carried by these wings that may also lead to electron acceleration to keV energy or more (Hess et al. *2007*). In the limit where $V_A \gg V_{flow}$, Alfvén wings are simply the magnetic flux tube connecting the obstacle to the source of the flow.

If the flow is itself magnetized enough, the conditions (accelerated electron distributions, plasma density, and magnetic field amplitude) at the footprints of the perturbed field lines can be favorable for the production of radio emission, in addition to that produced near the magnetized obstacle.

In the solar system, this does not happen in the solar wind, which is weakly magnetized and super-Alfvénic at all planetary orbits, but it does happen in the interaction of the magnetized satellite Ganymede with the rotating magnetic field of its parent planet Jupiter (Kivelson et al. *2004*), which leads to magnetic reconnection, Alfvén wings, electron acceleration, and radio emission generation near the footprints in Jupiter's ionosphere of the Ganymede flux tube – and also marginally near Ganymede (Kivelson et al. *2024*; Bonfond and Zarka *2024*). A similar interaction should exist between magnetized hot jupiters, orbiting close enough to their parent star to be in the region where the wind is still sub-Alfvénic. If the magnetic field amplitude at the stellar surface is strong enough (typically ≥30–100 times that of the Sun), radio emission should be produced not only in the exoplanet's auroral regions but also near the footprints on the stellar surface of the magnetic field lines connecting the hot jupiter to the star (Zarka *2006,2007*).

Note that in the Ganymede-Jupiter interaction, the Alfvén wings are connected to Jupiter roughly along a meridian plane perpendicular to $\mathbf{V}_{flow}$ and $\mathbf{B}_{flow}$ near Ganymede. The interaction between Jupiter's magnetic field and Ganymede is most often sub-Alfvénic, but sometimes super-Alfvénic (Kivelson et al. *2004*), but in both cases the Alfvénic perturbation can reach Jupiter's ionosphere on both ends of Ganymede's flux tube. By contrast, at most one Alfvén wing of a hot jupiter may travel upstream to the stellar surface (except if the planet lies in the closed magnetic field lines region (Preusse *2006*)) and this, only if the wind is sub-Alfvénic in the star's frame near the obstacle.

The transition between a super- and a sub-Alfvénic flow has been actually observed and simulated at Earth (Chané et al. *2012*, *2015*), which suggests that the envelope of a magnetosphere (the magnetopause) can be considered as the limiting case of the planet's Alfvén wings in a super-Alfvénic flow.

A particular case of the interaction between an exoplanet and a strongly magnetized wind concerns planets orbiting pulsars. Pulsar winds (beyond the light cylinder) are expected to be relativistic, so that both Alfvén wings of a planetary obstacle are carried away by the flow, independent of the magnetization of the obstacle (Mottez and Heyvaerts *2011,* Mottez et al. *2020*). It has been proposed that the plasma conditions in these wings are favorable to the development of an instability causing radio emission generation, and that relativistic focusing of the produced emission might explain the elusive fast radio bursts detected at cosmological distances (Mottez and Zarka *2014*).

## Unmagnetized Obstacle

When the obstacle is not magnetized, but is conductive (interior or ionosphere), its interaction with the plasma flow will lead to the development of an induced magnetosphere via the (temporary) pile-up of the wind's magnetic field on the obstacle's nose, caused by the slow large-scale perpendicular diffusion of plasma across magnetic fields. Conductivity is high in the ionospheres of Venus, Mars, Titan or comets close to the Sun, and possibly in the interior of metallic asteroids.

When the flow is super-Alfvénic, a bow shock again forms upstream of the obstacle, behind which magnetic pile-up occurs. Similar to the case of a magnetized obstacle, Alfvén waves cannot travel back to the source of the flow. Such induced magnetospheres are expected to exist for unmagnetized exoplanets orbiting within a magnetized super-Alfvénic stellar wind. The absence of a large-scale magnetic field of the obstacle prevents the focusing of accelerated electrons toward auroral regions and thus causes the absence of any intense radio emission near the obstacle for these systems.

In Jupiter's magnetosphere, the interaction of moon Callisto with the rotating Jovian field is in the super-Alfvénic regime. As noted for Ganymede, the particular geometry of this interaction does not prevent the Alfvénic perturbation to reach Jupiter's ionosphere on both ends of Callisto's flux tube, but no radio emission induced by Callisto has been detected so far.

When the flow is sub-Alfvénic, Alfvén wings similar to those described above develop in the flow past the obstacle, without any shock. The difference with the previous case is that the Alfvénic perturbations are not caused by magnetic reconnection but by the deviation of the flow (and its embedded magnetic field) by the conductive obstacle. Because they are driven only by the relative motion between the obstacle and the flow, these Alfvén wings are said to result from a "Unipolar Inductor" interaction (the theory was initially developed for artificial satellites in the Earth's ionosphere (Drell et al. *1965*)). The first and most famous example of this interaction is that of Io with the strong rotating Jovian magnetic field (Neubauer *1980*), which is now known to apply also to the Europa-Jupiter interaction (Jacome et al. *2022*). The conductivity of the obstacle is ensured by the existence of a volcanic atmosphere (and thus an ionosphere) around Io, and a subsurface ocean at Europa (Schilling et al. *2007*). Electrons accelerated in the Alfvén wings follow Jovian magnetic field lines down to the Jovian ionosphere. Above the northern and southern magnetic footprints of these two Galilean moons, the plasma conditions are favorable for the production of intense radio emission (Bigg *1964*; Louis et al. *2017*; Zarka et al. *2018*). Similar interactions may exist between some satellites of Saturn (e.g. Enceladus) and the planet's magnetic field, but because Saturn's magnetic field is much weaker than Jupiter's, the energy budget (see below) is not expected to give rise to strong satellite-induced radio emissions at Saturn, and indeed none has been observed so far.

This type of interaction is expected to take place in star-planet systems when the stellar wind is sub-Alfvénic (which should be the case for many hot jupiters) and the planet is unmagnetized. If the star is weakly (Sun-like) magnetized, then no radio emission is expected similar to the Saturn-satellite case. If the star is strongly magnetized ($\geq$30–100 times the Sun), then the star-planet system is a giant analogue of the Io-Jupiter (or Europa-Jupiter) system and strong radio emission is expected to be generated near one or both footprints of the magnetic flux tube connecting the exoplanet to the stellar surface (Zarka *2006, 2007*).

Willes and Wu (*2004*, *2005*) studied the particular case of terrestrial planets in close orbits around magnetic white dwarf stars, acting as a unipolar inductor, and generating radio emission with large flux densities. These systems could be remnants of main sequence stars with a planet that has survived the stellar expansion phase and has settled again on a stable orbit.

As mentioned above, the case of a conductive obstacle (planet, asteroid, dwarf companion) orbiting in a pulsar wind is the relativistic analogue of the Io-Jupiter interaction, for which Mottez and Heyvaerts (*2011*) have shown that although close to c, $V_{flow}$ is still likely to be lower than $V_A$, thus the interaction is expected to be analog to a unipolar inductor. Mottez et al. (*2020*) further noted that even in the super-Alfvénic regime, a conductive obstacle without an atmosphere (i.e. a metallic asteroid) will still develop Alfvén wings, within which an instability may develop and cause radio emission generation.

Finally, let us remind that when the obstacle is not only unmagnetized but also insulating (or when the flow is unmagnetized), it only absorbs the flow that impacts it, creating a plasma cavity in its wake, which is progressively refilled due to charged particle motion along the magnetic field lines permeating the flow. This is the case for the Earth's Moon (and probably also some rocky asteroids) in the solar wind. No bow shock nor any radio emission is produced in that configuration.

Table 1 lists all flow – obstacle interactions described above in a synthetic way, extrapolate them to star-planet interactions and predict their radio signatures.

**Table 1**

Possible types of flow-obstacle interactions, examples, radio signatures, and extrapolation to star-planet plasma interactions

| Obstacle | Flow Sub/Super Alfvénic | Flow B | Known Examples in the solar system<br><br>*Expected extrapolations to star-exoplanet systems* | Observed CMI radio signature<br><br>*Expected CMI radio signature* |
|---|---|---|---|---|
| B | Super | Weak | SoW-MS (magnetized planet : Mercury, Earth, Jupiter, Saturn, Uranus, Neptune), shock, reconnection, no AW<br>*StW-exoMS, shock, reconnection, no AW* | Planetary aurora<br><br>*Exoplanetary aurora* |
| B | Super | Strong | Jupiter-Ganymede[a], shock ?, reconnection + AW to Jupiter<br><br>*StW-exoMS, shock, reconnection, no AW upward the flow* | Ganymede aurora ?[b]<br>Induced emission near ionospheric footprints<br>*Exoplanetary aurora* |
| B | Sub | Weak | SoW-MS sometimes at Earth[c], no shock, reconnection + AW<br><br>*StW-exoMS, no shock, reconnection + AW* | Planetary aurora<br>-[c]<br><br>*Exoplanetary aurora*<br>*SPI* |
| B | Sub | Strong | Jupiter-Ganymede[a], no shock, reconnection + AW<br><br>*StW-exoMS, no shock, reconnection + AW*<br>*PW-exoMS, no shock, reconnection + AW* | Ganymede aurora ?[b]<br>Induced emission near ionospheric footprints<br>*Exoplanetary aurora*<br>*SPI*<br>*Pulsar planet aurora[d]*<br>*Instability in AW → FRB ?[e]* |
| B |  | None | Viscous-driven MS convection ?[f]<br>*Viscous-driven exoMS convection ?* | Planetary aurora ?<br>*Exoplanetary Aurora ?* |
| No B | Super | Weak | SW-conductive unmagnetized obstacle (Venus, Mars, Titan, metallic asteroid, comet close to the Sun), shock, induced MS, no AW<br>*StW-unmagnetized planet, shock, induced MS, no AW* | -<br><br><br>- |
| No B | Super | Strong | Jupiter-Callisto, shock ?, induced MS + AW to Jupiter<br><br>*StW-unmagnetized planet, shock, induced MS, no AW upward the flow*<br>*PW-asteroid, no shock, induced MS, AW* | -<br>Induced emission near ionospheric footprints ?[g]<br>*None or along Magnetopause ?*<br><br>*Instability in AW → FRB ?[h]* |
| No B | Sub | Weak | Saturn-conductive moon, no shock, AW<br><br>*StW-unmagnetized planet, no shock, induced MS + AW* | Induced emission near ionospheric footprints ?[g]<br>-<br>*SPI* |
| No B | Sub | Strong | Jupiter-Io, Jupiter-Europa, no shock, AW<br><br>*StW-unmagnetized planet, no shock, induced MS + AW*<br>*PW-asteroid, no shock, induced MS, AW* | Induced emission near ionospheric footprints<br>-<br>*SPI*<br>*Instability in AW → FRB ?[h]* |
| No B | - | None | Plasma wake | - |

Abbreviations : B = magnetic field ; SoW = solar wind ; MS = magnetosphere ; AW = Alfvén waves ; StW = stellar wind ; exoMS = exoplanet's magnetosphere ; SPI = Star-planet plasma interaction with radio emission near the star induced by the stellar wind-exoplanet sub-Alfvénic interaction, if $f_{pe}/f_{ce}$ allows ; PW = Pulsar wind

Notes : [a] Ganymede-Jupiter interaction is mostly sub-Alfvénic, except when Ganymede crosses Jupiter's current sheet, where the Alfvén speed is lower ; [b] weak low-frequency upper hybrid emission has been observed near Ganymede (Kivelson et al. *2024*) ; [c] rare cases studied by Chané et al. (*2012*, *2015*), with too limited duration and too far from the Sun for generating a SPI induced emission ; [d] cf. Mishra et al. (*2023*) ; [e] cf. Mottez and Zarka (*2014*) ; [f] cf. (Axford and Hines, *1974*) ; [g] weak if any, undetected ; [h] cf. Mottez et al. (*2020*), Voisin et al. (*2021*)

# Dissipated Power and Radio Signatures

From Table 1 it appears clearly that the generation of intense radio emission from the system requires either a magnetized obstacle or a sub-Alfvénic interaction with a strongly magnetized flow. This is systematically verified in the solar system. This can be understood considering that the most efficient and ubiquitous mechanism for generating magnetospheric (auroral and satellite-planet) radio emissions is the Cyclotron Maser Instability (CMI – Zarka 1998). The CMI is a wave-particle instability that directly converts the perpendicular energy of electrons gyrating in a large-scale magnetic field into electromagnetic (radio) energy (Wu and Lee 1979). It has been demonstrated that up to a few percent of the electrons' energy can be converted into radio waves, if three conditions are met: the presence of a strong magnetic field, of accelerated electrons with keV (or tens of keV) energies, and a relatively depleted and strongly magnetized plasma at the source (quantified by $f_{pe}/f_{ce} \ll 1$, with $f_{pe}$ the plasma frequency and $f_{ce}$ the electron cyclotron frequency in the source). CMI radio emission is strongly circularly polarized, very anisotropic, and produced at frequencies f close to $f_{ce}$ in the source. As $f_{ce}$ varies continuously along magnetic field lines, a broadband continuum emission is expected (below a few tens of MHz for Jupiter's radio emissions, and ≤1 MHz for other planetary radio emissions in the solar system (Zarka 1998)). The CMI is by far the most efficient radio emission mechanism, and it dominates all other ones (such as beam-plasma instabilities) when the conditions at the source are met. This is why the auroral radio emissions of solar system planets – especially that of Jupiter – are as bright as solar radio bursts (that are mostly not produced by the CMI due to $f_{pe}/f_{ce} > 1$ in the solar corona except at specific locations, e.g., at footprints of strong magnetic loops (Morosan et al. 2016)).

CMI intensity depends on the specific features present in the keV electron distributions in the source, and is consequently not predictable from first principles. Instead, an empirical way was found to predict emitted radio powers and flux densities, by comparing it to the electromagnetic power involved in the corresponding flow-obstacle interactions (Zarka et al. 2001; Zarka 2007). It was found that in all cases the emitted radio power $P_r$ follows the relation:

$$P_r = \beta \, (V_{flow} B_\perp^2/\mu_o) \, \pi R_{obs}^2 \qquad (1)$$

with $B_\perp$ the flow's magnetic field component perpendicular to the flow direction in the obstacle's frame, and an efficiency factor $\beta = (2-10) \times 10^{-3}$. This very simple relation was found to unify all solar system magnetospheric auroral emissions and satellite-induced radio emissions at Jupiter, which is not the case for a relation based on the dynamic pressure of the flow (Zarka et al. 2018). It is illustrated in Fig. 1. This relation will be called hereafter the radio-magnetic scaling law, or RMSL (rather than the radio Bode's law, or RBL, initially proposed for a radio-dynamic pressure relation (Desch and Kaiser 1984)).

The empirical RMSL (1) led Zarka et al. (2001) and Zarka (2007) to propose that the electromagnetic power $P_d$ dissipated in all known flow-obstacle interactions can be expressed as:

$$P_d = \varepsilon \, (V_{flow} B_\perp^2/\mu_o) \, \pi R_{obs}^2 \qquad (2)$$

Equation (2) is thus simply the fraction $\varepsilon$ of the Poynting flux (or magnetic energy flux) $V_{flow} B_\perp^2/\mu_o$ convected on the obstacle of cross section $\pi R_{obs}^2$. The flow's dynamic (and thermal) pressure is contained in the expression of $R_{obs}$ for a magnetospheric obstacle, that is taken as $R_{obs} = 1.5 \, R_{MP}$ with $R_{MP}$ the magnetopause radius resulting from pressure balance inside and outside the magnetosphere:

$$R_{MP} = [\, 2 \, B_p^2 \, / \, (\mu_o n_{sw} m_p V_{sw}^2 + B_{sw}^2/2) \,]^{1/6} \qquad (3)$$

with $B_p$ the planetary equatorial surface field, $m_p$ the proton's mass, and $n_{sw}$, $V_{sw}$ and $B_{sw}$ the number density, velocity and magnetic field of the solar wind flow (see Varela et al. (*2022*) for the general expression including thermal pressure).

Theoretical considerations suggest that the efficiency of the dissipation ε is comprised between $M_A$ (in the sub-Alfvénic case) and 1 (Zarka et al. *2001*; Zarka *2007*). Measured efficiencies for the Earth's magnetosphere and satellite-Jupiter interactions are ε = 0.2 ± 0.1. Hence, as $P_r = β/ε\ P_d$, one deduces that a fraction 1–5% of $P_d$ is converted into radio waves. The value 1% is now the canonical value used in all theoretical predictions. Observed radio signatures can thus help to evaluate the power dissipated in an interaction, and deduce physical parameters of the observed flow-obstacle system.

For example, expression (2) and the variation as a function of the distance to the Sun of $B_⊥$ in the frame of a planet in Keplerian orbit, allowed Zarka et al. (*2001*) and Zarka (*2007*) to identify a corridor at 0.17 AU where the flow and the magnetic field are nearly aligned with each other and thus no or very little Poynting flux is generated in the interaction. This corridor should exist in all stellar winds near 0.1–0.3 AU (Saur et al. *2013*).

The RMSL also led Zarka (*2006*, *2007*) to note that in the case of HD 179949, the system for which the first optical evidence of a star-planet interaction was detected by Shkolnik et al. (*2003*, *2004*), the demand put on the dissipated power by the energetics of the optical signature required a stellar and/or a planetary magnetic of several tens of Gauss. This prediction was supported by Saur et al. (*2013*), and confirmed by Yadav and Thorngren (*2017*) and Cauley et al. (*2019*) who concluded to a surface polar planetary magnetic field of 90±20 G.

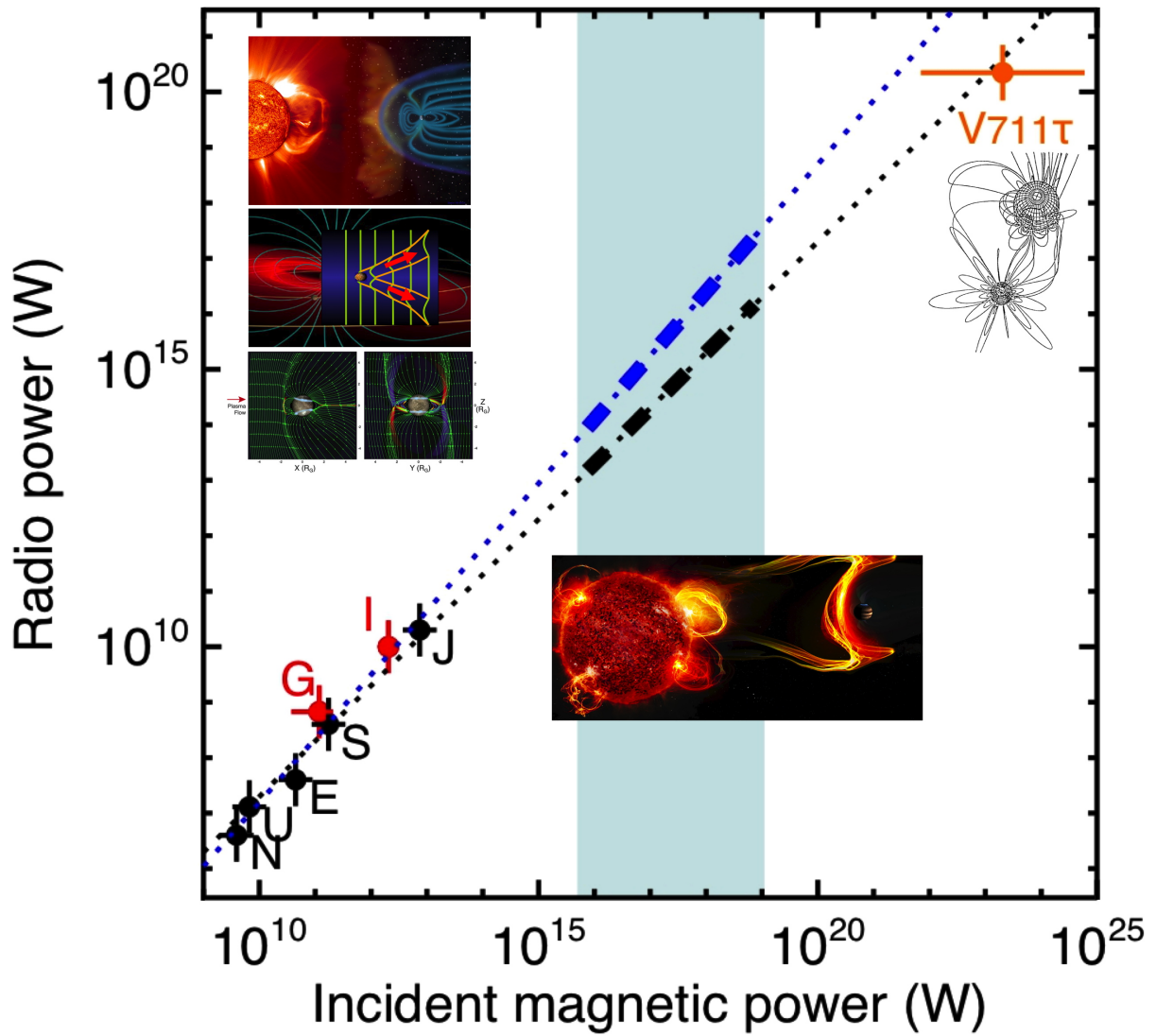

**Fig. 1**

Radio-magnetic scaling law relating magnetospheric auroral (E, J, S, U, N = Earth, Jupiter, Saturn, Uranus, and Neptune) and satellite-induced (I, G = Io, Ganymede) radio power to incident Poynting flux of the plasma flow on the obstacle. *Black dotted line* has slope 1, emphasizing the proportionality between ordinates and abscissae, with a coefficient $2 \times 10^{-3}$. The best fit has a slope of 1.15 (blue dotted line). The *thick dashed lines* extrapolate the radio-magnetic scaling law to the parameter ranges for hot jupiters (*shaded region*) where the intercepted Poynting flux can reach $10^3$–$10^6$ times that intercepted by Jupiter (taking into account increased magnetospheric compression and a solar-like magnetic field of ~1 Gauss). Stronger stellar magnetic fields should lead to stronger radio emissions. The *orange dot* illustrates the case of the RS CVn magnetic binary V711 Tauri discussed in the text. *Insets sketch* the types of interaction (solar wind-magnetosphere, Jupiter-Io, Jupiter-Ganymede, and star-hot jupiter).

Several theoretical works attempted to explore the causes and efficiency of the conversion of the incident magnetic energy flux into the energy of accelerated electrons. Jardine and Cameron (*2008*) proposed that runaway electrons are accelerated by the electric field resulting from magnetic reconnection at the flow-obstacle interaction, but this does not seem to be a significant process for generating solar system radio emissions. Saur et al. (*2013*), computing analytically the magnetic energy fluxes dissipated in sub-Alfvénic interactions, and found a result about a factor ×2 above that of equation (2) (see. Callingham et al. (*2024*) for details), with dissipated energy fluxes $\geq 10^{19}$ W in star-planet interactions, to be compared with $10^{12}$ W per hemisphere in the Io-Jupiter case. Nichols and Milan (*2016*) modelled analytically the solar wind-Earth magnetosphere coupling transposed to a regime of parameters corresponding to hot Jupiters. They predicted radio powers up to $10^{15}$ W, increasing with the planetary magnetic field and decreasing with increasing orbital distance, as expected. Their predicted values range from 1.5 orders of magnitude below to 1 order of magnitude above the RMSL predictions, depending on the star's age (and thus the stellar wind strength) and the planet's orbital distance. For exoplanets very close to the star, the planetary magnetospheric convection induced by the stellar wind may "saturate", preventing to dissipate the total available incident Poynting flux.

MHD simulations also explored quantitatively the physics subtending the RMSL. For close-in exoplanets in the sub-Alfvénic region, Ip et al. (*2004*) and Strugarek et al. *(2015)* explored the large variability of the interaction with the relative topology of the planetary and interplanetary (or stellar wind) magnetic fields. In the super-Alfvénic case, dissipated magnetic energy fluxes and predicted radio emission power have been computed for the solar wind-Mercury interaction by Varela et al. (*2016*). $P_d$ and $P_r$ were found to be 3-6 times lower than the RMSL predictions, to increase with $B_{sw}$ as expected, but more surprisingly also with the solar wind dynamic pressure $P_{dyn}$. Part of the incident Poynting flux is dissipated at the dayside magnetopause, and part in the night side magnetotail.

The simulations of Varela et al. (*2018, 2022*) and Peña-Moñino et al. (*2024*) confirmed the higher efficiency of the RMSL compared to the radio-$P_{dyn}$ relation, the increase of $P_r$ with both $B_{sw}$ and $B_p$, and found a larger $P_r$ for $\mathbf{B}_{sw}$ // $\mathbf{B}_p$ (i.e. anti-aligned magnetic fields at the magnetopause nose, maximizing reconnection). The overall predictions for $P_r$ are within one order of magnitude of those from the RMSL. These simulations explored the large dependence of $P_d$ and $P_r$ on the planetary vs. stellar wind magnetic topology: in sub-Alfvénic regime, a variability of ×10 of $P_d$ and $P_r$ with the orientation of $\mathbf{B}_{sw}$ was found, on the average two times lower than the RMSL, whereas in super-Alfvénic regime the variability is less (×2) but on the average ten times lower than the RMSL; this should in turn cause large time-variability in the exoplanetary radio signals). Finally, they confirmed the amplifying role of the stellar wind dynamic pressure $P_{dyn}$ : a large magnetospheric compression leads to larger magnetic energy dissipation and radio emission, up to ×100 following impact of a large CME (Coronal Mass Ejection).

It is rather surprising that the Jupiter dot on Fig. 1 lies on the RMSL line, because Jupiter's auroral radio emission mainly results from internal dynamics, namely the magnetosphere-ionosphere (M-I) coupling and corotation breakdown for the Iogenic plasma ejected centrifugally (Cowley and Bunce *2001*). This may be partly due to the fact that a fraction of the Jovian radio emission is controlled by the solar wind (Zarka and Genova *1983*), and be partly coincidental, the parameters of M-I coupling at Jupiter being what they are. Nichols (*2011*) extrapolated the study of M-I coupling to exoplanetary Jupiter-like magnetospheres and found that the radio output can reach $10^{16}$ W for fast rotating magnetospheres of strongly magnetized planets orbiting at a few AU from their parent star, with a strong stellar X-UV luminosity being a favourable factor maximizing the M-I coupling. But also note that M-I coupling can be qualitatively understood as a limiting case of flow-obstacle interaction where the obstacle is the inertia of the plasma that resists displacement in corotation with planetary magnetic field lines. As a consequence, magnetic field-aligned-currents arise to communicate the torque between the magnetosphere and the ionosphere. These currents are generally larger than that which can be carried by unaccelerated magnetospheric electrons, so that field-aligned potential drops appear, that cause electron acceleration to keV energies, and ultimately radio emissions. These processes are suspected to play a role in the generation of radio emission from cool dwarf stars, via the coupling of the stellar magnetic field with its plasma environment (Hallinan et al. *2015*; Callingham et al. *2021a*).

To test the extrapolation of the radio-magnetic scaling law toward high Poynting fluxes and high radio powers, Zarka (*2010*) analyzed the literature on radio emission from magnetized binary stars. They found that for the RS CVn stellar system V711 Tauri, measured radio flux densities (0.1–1 Jy – Budding et al. *1998*; Richards et al. *2003*) and distance (29 pc), and estimated magnetic fields (10–30 G at the interaction region) and bandwidth (≥8 GHz) allow to estimate that $10^{-4} \leq \beta \leq 10^{-2}$ (see the detailed calculation in (Mottez and Zarka *2014*) ; the lower end of this range may be explained by a radio emission from V711 Tauri actually produced by gyrosynchrotron rather than CMI, hence with a lower efficiency). This good agreement with the solar system value of β ($2 \times 10^{-3}$) suggests that the RMSL holds – at least approximately – for 10 orders of magnitude above the range of solar system planets.

## Interpreting Radio Signatures

CMI radio emissions are expected to occur in the spectral range below a few tens of MHz, unless exoplanets much more strongly magnetized than Jupiter exist, which is not excluded (Reiners and Christensen *2010*; Yadav and Thorngren *2017*). For close-in exoplanets, that are probably spin-orbit locked, the relatively long rotation period (equal to the orbital period) may lead to a decay of the planetary dynamo and thus of the magnetic field (Sanchez-Lavega *2004*), and the expanded ionospheres of hot Jupiters may prevent CMI radio emission production or escape (Weber et al. *2017*). But some of the star-planet interactions described in Table 1 should lead to radio emissions at star-planet flux tube footprints, i.e., at cyclotron frequencies governed by the stellar magnetic field. These emissions might reach hundreds of MHz or more.

At these frequencies, the angular resolution of milli-arcseconds that would be necessary to separate the radio emission of an exoplanet from that of its parent star will not be available in the foreseeable future. Emission will thus be detected as polarized point sources in surveys, and the most interesting information will be obtained through dynamic spectra, i.e., measurements of the intensity and polarization as a function of time and frequency. With these data, discrimination between the stellar (coronal) and exoplanetary or exoplanet-induced CMI radio emissions can be done via the presence of circular polarization, time-frequency structure of the signal and especially its modulation at the exoplanet's orbital period or, in some cases, at the synodic period between this orbital period and the star's rotation (Louis et al. *in prep.*).

As the CMI theory provides today a well-understood, quantitative framework for understanding radio emissions properties, radio intensity and polarization measured as a function of time and frequency will provide powerful diagnostics of the plasma environments and processes at work in star-planet interactions. This interpretation framework is at the origin of the code ExPRES

(exoplanetary and planetary radio emissions simulator), developed and well tested for Jupiter's radio emissions (Hess et al. *2008*), and finalized and made public by Louis et al. (*2019*). Using this code, simulations by Hess and Zarka (*2011*) have shown that such measurements will give access to the type of interaction at work (exoplanet's auroral emission or exoplanet-induced emission in the stellar field), its energetics, the exoplanet's magnetic field intensity and tilt (if emission comes from the exoplanet's magnetosphere), the planetary rotation and revolution periods (effectively testing tidal spin-orbit synchronization), and the orbit inclination (resolving the ambiguity on the planet's mass). Figure 2 illustrates how these parameters can be deduced from the observed dynamic spectra. Temporal modulations different from the planetary rotation and orbital periods could additionally reveal the presence of moons or the signature of the stellar wind activity. Most of these parameters, especially those concerning the exoplanet's magnetic field and thus its interior structure, are very difficult or impossible to determine by other means than radio measurements.

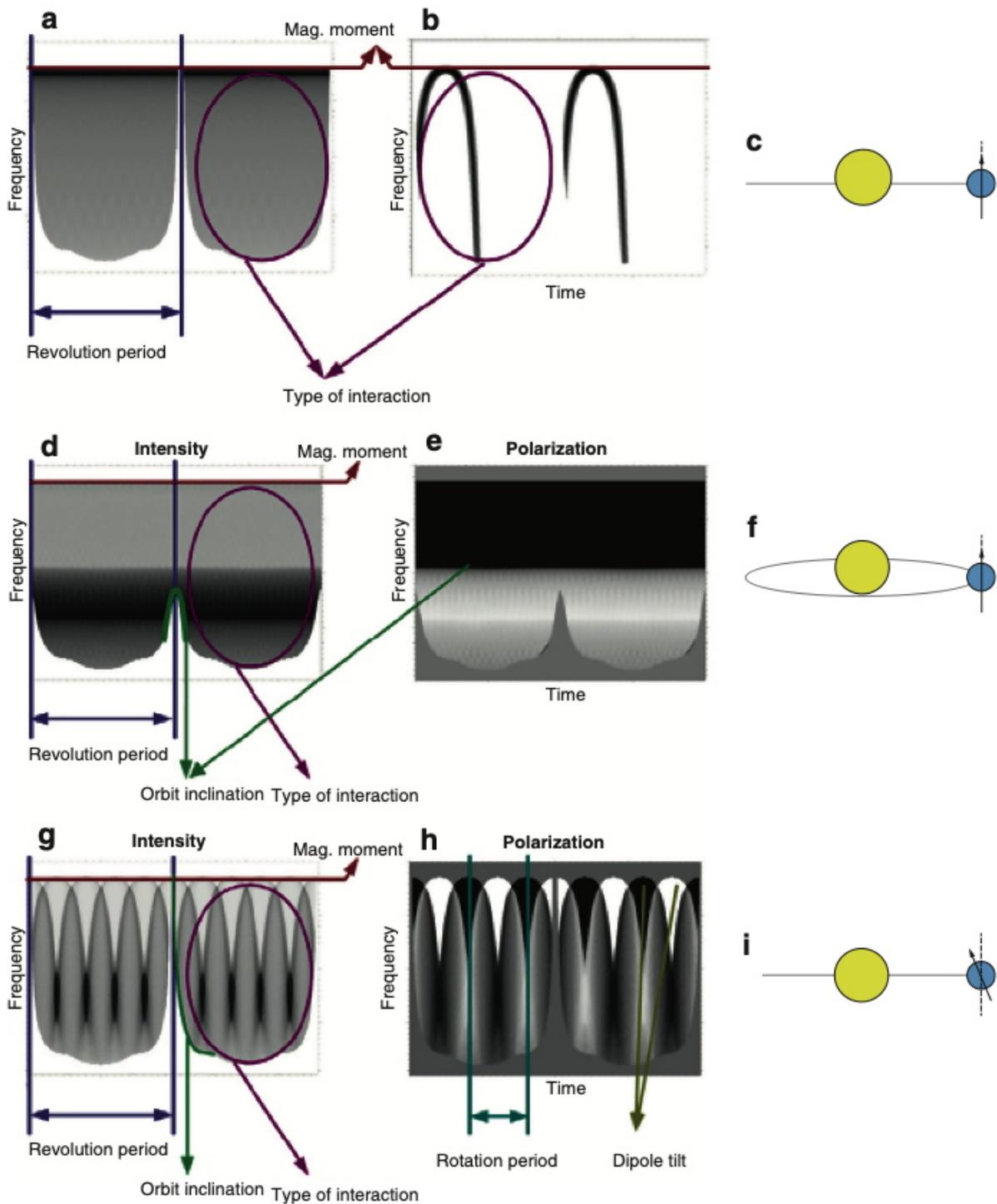

**Fig. 2**

Simulated dynamic spectra in intensity (**a**), (**b**), (**d**), (**g**) and circular polarization (**e**), (**h**) and associated parameters of the system that can be determined: (**a**), (**b**), (**c**) exoplanetary magnetic field aligned with the rotation axis and 0° orbit inclination, for emission (**a**) of a full exoplanet's auroral oval and (**b**) an auroral active sector fixed in local time. (**d**), (**e**), (**f**) aligned exoplanetary magnetic field and 15° orbit inclination, for emission of a full auroral oval. (**g**), (**h**), (**i**) exoplanetary magnetic field tilted by 15° and 0° orbit inclination, for emission of a full auroral oval (From Hess and Zarka ([2011](#)))

It must be noted that the existence of an exoplanet's magnetosphere is an essential ingredient in favour of planet habitability: it ensures shielding of the planet's atmosphere and surface, preventing $O_3$ destruction by cosmic rays bombardment as well as atmospheric erosion by the stellar wind or CMEs (Grieβmeier et al. *2004*, *2005*); it may also limit atmospheric escape but this is debated (Gronoff et al. *2020*). Simulations by Varela et al. (*2016*, *2018*, *2022*) may feed Bayesian models able to interpret the radio emission intensity and its time variability in terms of planet shielding efficiency and magnetopause standoff distance: moderate emission intensity and time variability seem favourable for habitability (whereas a large variability may reveal strong magnetic reconnection with the interplanetary magnetic field, and intense weakly variable radio emission is the sign of a compressed magnetosphere, both unfavourable to habitability). Long time series of radio measurements are thus crucial for detection and characterization of the targets.

The code ExPRES was also recently used to derive, from observations at ~1 GHz with the giant radiotelescope FAST of radio bursts from the active star AD Leonis, the location and energy of the electrons producing the radio bursts (Zarka et al., *2024*).

Radio detection may eventually evolve as an independent discovery tool, complementary to radial velocities or transit measurements because it is more adapted to finding planets (giant or terrestrial) around and interacting with active, magnetic or variable stars.

# Detection Prospects

In the past few years, circularly polarized radio emissions have been detected from several stars – mostly M dwarfs – at frequencies around 1 GHz and attributed to the CMI (Osten et al. *2006*, *2008*; Hallinan et al. *2007*, *2015*; Villadsen and Hallinan *2019*; Zic et al. *2019*; Bastian et al. *2022*; Zhang et al. *2023*). The LOFAR imaging survey at wavelength λ~2 m extended these detections to frequencies ~150 MHz (Callingham et al. *2021a,b, 2023*), including one stellar system with a known exoplanet (GJ 625), but without being able to conclude about the origin of the radio emission: stellar, planetary or SPI ? Time-variable emission was also detected with LOFAR from the quiet dwarf GJ 1151 and attributed to SPI with a putative planet that has not been confirmed (Vedantham et al. *2020*). Radio bursts were tentatively detected with LOFAR in dynamic spectrum mode at 14-21 MHz (Turner et al. *2021*) but were not confirmed by follow-up observations (Turner et al. *2024*), which may be due to the intrinsic emission variability (Elekes and Saur *2023*). At higher frequencies (1-4 GHz), circularly polarized bursts were detected from the systems of Proxima Centauri (Perez-Torres et al. *2021*) and YZ Ceti (Pineda and Villadsen *2023*), again without being able to conclude about the origin of the radio emission. New methods allowing to build dynamic spectra from imaging data are now being used for re-analyzing LOFAR data (Tasse et al. *in prep.*) and analyzing NenuFAR ones (Zarka et al. *2020*).

# Conclusion

Star-planet plasma interactions are expected to be common, occurring when the exoplanet is magnetized or located in the sub-Alfvénic region of the stellar wind. The proposed radio-magnetic scaling law summarized by Eqs. ([1](#)) and ([2](#)) is of course an approximation, but its principle is now widely accepted and used. Various analytical works and numerical simulations (Nichols [2011](#); Saur et al. [2013](#); Nichols and Milan [2016](#); Varela et al. [2016, 2018, 2022](#); Peña-Moñino et al. [2024](#)) suggest that deviations of one order of magnitude are expected (up to a factor ×30 in extreme cases), that can be incorporated in an uncertainty on $\beta$ ($10^{-4} \leq \beta \leq 10^{-2}$), over a range of validity that spans >10 orders of magnitude in flux density above the range of solar system planets. The RMSL can thus be used for predicting and interpreting radio flux densities from exoplanets and star-planet interacting systems. The code PALANTIR (Mauduit et al. [2023](#)) has been built to feed the RMSL with measured or inferred parameters of star-planet systems and predict frequencies and flux densities emitted in radio. The RMSL predicts that radio emissions up to $10^{5-6}$ times more intense than Jupiter's should exist, especially in hot jupiter systems. Even if hot jupiters are weakly or not magnetized due to slow sidereal rotation, the possibility of sub-Alfvénic interaction with a strongly magnetized parent star via Alfvén waves offers serious perspectives for radio detection, at frequencies up to hundreds of MHz, larger than those expected for magnetospheric emissions (that are below a few tens of MHz). Stellar magnetic field measurements via Zeeman Doppler spectropolarimetry (e.g., Vidotto et [al.](#) 2014) is a crucial information for selecting candidates for radio emission.

Radio emissions $10^{3-4}$ times stronger than Jupiter's should be detectable at stellar distances (pc to tens of pc) with existing large low-frequency radiotelescopes (LOFAR in Europe, NenuFAR in France), and SKA will make detection possible for emissions $10^{1-2}$ times stronger than Jupiter's (Zarka et al. [2015](#)). SKA will perform large surveys, and the observing parameters adapted to the radio detection of exoplanets and star-planet interactions are also adapted to the study of stellar coronal bursts or CMI emission from cool stars. Pulsar planets may also produce detectable radio flux (Mishra et al. [2023](#)).

Based on the spatial in situ exploration of planetary magnetospheres in our solar system, a reliable interpretation frame is ready (Hess and Zarka [2011](#)), waiting for the first unambiguous detection (only tentative detections have been made until now - see above). Confirmed radio detections should occur soon, which will open the new field of comparative exo-magnetospheric physics, exo-space weather and allow us to probe flow-obstacle interactions in various regimes and over a large range of parameters (star-planet distance, stellar magnetic field and wind strength, stellar X-UV flux, planetary magnetic field, rotation, orbit inclination, etc.). Radio measurements (flux density, variability) are a unique window on exoplanetary magnetic fields, will provide constraints on exoplanet habitability, and will allow us to infer the energy dissipation involved in the corresponding star-planet interactions and thus to refine the radio-magnetic scaling law.

# Cross-References

**Acknowledgments**

PZ acknowledges funding from the programs PNP, PNST, PNPS, and AS SKA-LOFAR of CNRS/INSU, and from the ERC No 101020459—Exoradio.

# References

Axford WI, Hines CO (1974) A unifying thepry of high-latitude geophysical phenomena and geomagnetic storms. In: Hines CO (ed) The upper atmosphere in motion. Geophysical Monograph Series, Washington D.C., AGU, n°18:936-967

Bagenal F (2001) Planetary magnetospheres. In: Murdin P (ed) Encyclopedia of astronomy and astrophysics. IOP Publishing, Bristol. article 2329

Bastian TS, Cotton WD, Hallinan G (2022) Radio Emission from UV Cet: Auroral Emission from a Stellar Magnetosphere. Astrophys J 935:99 (16pp)

Bigg EK (1964) Influence of the satellite Io on Jupiter's decametric emission. Nature 203:1008–1010

Bonfond B, Zarka P (2024) Electrodynamic Coupling between Ganymede and the Jovian Ionosphere. In: Volwerk M, McGrath M, Jia X et al (eds) Ganymede. Cambridge University Press, Cambridge, Chapter 3.6, in press

Budding E, Slee OB, Jones K (1998) Further discussion of binary star radio survey data. PASA 15:183–188

Callingham JR, Pope BJS, Feinstein AD et al (202a1) Low-frequency monitoring of flare star CR Draconis: long-term electron-cyclotron maser emission. Astron Astrophys 648:A13

Callingham JR, Vedantham HK, Shimwell TW et al (2021b) The population of M dwarfs observed at low radio frequencies. Nature Astronomy 5:1233-1239

Callingham JR, Shimwell TW, Vedantham HK et al (2023) V-LoTSS: The circularly polarised LOFAR Two-metre Sky Survey. Astron Astrophys 670:A124

Callingham JR, Pope BJS, Kavanagh RD et al (2024) Radio Signatures of Star-Planet Interactions, Exoplanets and Space Weather. Nature Astronomy in press

Cauley PW, Shkolnik EL, Llama J et al (2019) Magnetic field strengths of hot Jupiters from signals of star–planet interactions. Nature Astronomy 3:1128-1134

Chané E, Saur J, Neubauer FM et al (2012) Observational evidence of Alfvén wings at the Earth. J Geophys Res 117:A09217

Chané E, Raeder J, Saur J et al (2015) Simulations of the Earth's magnetosphere embedded in sub-Alfvénic solar wind on 24 and 25 May 2002. J Geophys Res 120:8517–8528

Cowley SWH, Bunce EJ (2001) Origin of the main auroral oval in Jupiter's coupled magnetosphere–ionosphere system. Planet Space Sci 49:1067-1088

Cuntz M, Saar SH, Muzeliak ZE (2000) On stellar activity enhancement due to interactions with extrasolar giant planets. Astrophys J 533:L151–L154

Desch MD, Kaiser ML (1984) Predictions for Uranus from a radiometric Bode's law. Nature 310:755-757

Drell SD, Foley HM, Ruderman MA (1965) Drag and propulsion of large satellites in the ionosphere: an Alfvén propulsion engine in space. J Geophys Res 70(13):3131–3145

Dungey JW (1961) Interplanetary magnetic field and the auroral zones. Phys Rev Lett 6(2):47-48

Elekes F, Saur J (2023) Space environment and magnetospheric Poynting fluxes of the exoplanet $\tau$ Boötis b. Astron Astrophys 671:A133

Encrenaz T, Bibring JP, Blanc M et al (2004) The solar system, 3rd edn. A&A Library, Springer, Berlin. http://www.springer.com/in/book/9783540002413

Grießmeier JM, Stadelmann A, Penz T et al (2004) The effect of tidal locking on the magnetospheric and atmospheric evolution of "hot jupiters". Astron Astrophys 425:753–762

Grießmeier JM, Motschmann U, Mann G, Rucker HO (2005) The influence of stellar wind conditions on the detectability of planetary radio emissions. Astron Astrophys 437:717–726

Gronoff G, Arras P, Baraka S, et al (2020) Atmospheric escape processes and planetary atmospheric evolution. J Geophys Res 125:e2019JA027639

Hallinan G, Bourke S, Lane C et al (2007) Periodic bursts of coherent radio emission from an ultracool dwarf. Astrophys J 663:L25–L28

Hallinan G, Littlefair SP, Cotter G et al (2015) Magnetospherically driven optical and radio aurorae at the end of the stellar main sequence. Nature 523(7562):568–571

Hess S, Mottes F, Zarka P (2007) Jovian S burst generation by Alfvén waves. J Geophys Res 112:A11212

Hess SLG, Zarka P (2011) Modeling the radio signature of the orbital parameters, rotation, and magnetic field of exoplanets. Astron Astrophys 531:A29

Hess S, Cecconi B, Zarka P (2008) Modeling of Io-Jupiter decameter arcs, emission beaming and energy source. Geophys Res Lett 35:L13107

Ip WH, Kopp A, Hu JH (2004) On the star–magnetosphere interaction of close-in exoplanets. Astrophys J 602:L53–L56

Jacome HRP, Marques MS, Zarka P et al (2022) Search for Jovian DAM emissions induced by Europa on the extensive Nançay Decameter Array's catalog. Astron Astrophys 665:A67

Jardine M, Cameron AC (2008) Radio emission from exoplanets: the role of the stellar coronal density and magnetic field strength. Astron Astrophys 490:843–851

Kivelson MG, Bagenal F, Kurth WS et al (2004) Magnetospheric interactions with satellites. In: Bagenal F, McKinnon W, Dowling T (eds) Jupiter: the planet, satellites, and magnetosphere. Cambridge University Press, Cambridge, pp 513–536

Kivelson MG, Bagenal F, Jia X et al (2024) Ganymede: Its Magnetosphere and its Interaction with the Jovian Magnetosphere. In: Volwerk M, McGrath M, Jia X et al (eds) Ganymede. Cambridge University Press, Cambridge, Chapter 3.1, in press

Lepping RP (1986) Magnetic configuration of planetary obstacles. In: Comparative study of magnetospheric systems. Cepadues/CNES ed, Toulouse, pp 45–75

Louis CK, Lamy L, Zarka P, Cecconi B, Hess SLG (2017) Detection of Jupiter decametric emissions controlled by Europa and Ganymede with Voyager/PRA and Cassini/RPWS. J Geophys Res (in press)

Louis CK, Hess SLG, Cecconi B et al (2019) ExPRES: an Exoplanetary and Planetary Radio Emissions Simulator. Astron Astrophys 627:A30

Mauduit E, Grießmeier JM, Zarka P et al (2023) PALANTIR: an updated prediction tool for exoplanetary radio emissions. In: Louis CK, Jackman CM, Fischer G et al (eds) Planetary, Solar and Heliospheric Radio Emissions IX. DIAS, TCD:485–496

Mishra R, Čemeljić M, Varela J et al (2023) Auroras on Planets around Pulsars. Astrophys J Lett 959:L13 (6pp)

Morosan DE, Gallagher PT, Zucca P et al (2016) LOFAR tied-array imaging and spectroscopy of solar S bursts. Astron Astrophys 580:A65

Mottez F, Heyvaerts J (2011) Magnetic coupling of planets and small bodies with a pulsar wind. Astron Astrophys 532:A21

Mottez F, Zarka P (2014) Radio emissions from pulsar companions: a refutable explanation for galactic transients and fast radio bursts. Astron Astrophys 569:A86

Mottez F, Zarka P, Voisin G (2020) Repeating fast radio bursts caused by small bodies orbiting a pulsar or a magnetar. Astron Astrophys 644:A145

Neubauer FM (1980) Nonlinear standing Alfvén wave current system at Io: theory. J Geophys Res 85(A3):1171–1178

Nichols JD (2011) Magnetosphere–ionosphere coupling at Jupiter-like exoplanets with internal plasma sources: implications for detectability of auroral radio emissions. MNRAS 414:2125–2138

Nichols JD, Milan SE (2016) Stellar wind–magnetosphere interaction at exoplanets: computations of auroral radio powers. MNRAS 461:2353–2366

Peña-Moñino L, Pérez-Torres M, Varela J et al (2024) Magnetohydrodynamic simulations of the space weather in Proxima b: Habitability conditions and radio emission. Astron Astrophys in press

Pérez-Torres M, Gomez JF, Ortiz JL et al (2021) Monitoring the radio emission of Proxima Centauri. Astron Astrophys 645:A77

Pineda JS, Villadsen J (2023) Coherent radio bursts from known M-dwarf planet-host YZ Ceti. Nature Astronomy 7:569-578

Osten RA, Bastian TS (2006) Wide-band spectroscopy of two radio bursts on AD Leonis. Astrophys J 637:1016-1024

Osten RA, Bastian TS (2008) Ultrahigh time resolution observations of radio bursts on AD Leonis. Astrophys J 637:1078-1085

Preusse S, Kopp A, Büchner J, Motschmann U (2006) A magnetic communication scenario for hot jupiters. Astron Astrophys 460:317–322

Reiners A, Christensen UR (2010) A magnetic field evolution scenario for brown dwarfs and giant planets. Astron Astrophys 522:A13

Richards MT, Waltman EB, Ghigo FD, Richards DSP (2003) Statistical analysis of 5 year continuous radio flare data from β Persei, V711 Tauri, δ Librae, and Ux Arietis. Astrophys J Suppl Ser 147:337–361

Sanchez-Lavega A (2004) The magnetic field in giant extrasolar planets. Astrophys J 609:L87–L90

Saur J, Grambusch T, Duling S, Neubauer FM, Simon S (2013) Magnetic energy fluxes in sub-Alfvénic planet star and moon planet interactions. Astron Astrophys 552:A119

Schilling N, Neubauer FM, Saur J (2007) Time-varying interaction of Europa with the jovian magnetosphere: Constraints on the conductivity of Europa's subsurface ocean. Icarus 192:41–55

Shkolnik E, Walker GAH, Bohlender DA (2003) Evidence for planet-induced chromospheric activity on HD 179949. Astrophys J 597:1092–1096

Shkolnik E, Walker GAH, Bohlender DA (2004) Erratum: evidence for planet-induced chromospheric activity on HD 179949. Astrophys J 609:1197

Strugarek A, Brun AS, Matt SP, Reville V (2015) Magnetic games between a planet and its host star: the key role of topology. Astrophys J 815(2):111

Turner JD, Zarka P, Grießmeier JM et al (2021) The search for radio emission from the exoplanetary

systems 55 Cancri, ν Andromedae and τ Boötis using LOFAR beam-formed observation. Astron Astrophys 645:A59

Turner JD, Zarka P, Grießmeier JM et al (2024) Follow-up LOFAR observations of the τ Boötis exoplanetary system. Astron Astrophys in press

Varela J, Reville V, Brun AS, Pantellini F, Zarka P (2016) Radio emission in Mercury magnetosphere. Astron Astrophys 595:A69

Varela J, Réville V, Brun AS et al (2018) Effect of the exoplanet magnetic field topology on its magnetospheric radio emission. Astron Astrophys 616:A182

Varela J, Brun AS, Zarka P et al (2022) MHD study of extreme space weather conditions for exoplanets with Earth-like magnetospheres: On habitability conditions and radio-emission. Space Weather 20:e2022SW003164

Vedantham HK, Callingham JR, Shimwell TW et al (2020) Coherent radio emission from a quiescent red dwarf indicative of star–planet interaction. Nature Astronomy 4:577:583

Vidotto AA, Gregory SG, Jardine M et al (2014) Stellar magnetism: empirical trends with age and rotation. MNRAS 441:2361–2374

Villadsen J, Hallinan G (2019) Ultra-wideband Detection of 22 Coherent Radio Bursts on M Dwarfs. Astrophys J 871:214 (28pp)

Voisin G, Mottez F, Zarka P (2021) Periodic activity from fast radio burst FRB180916 explained in the frame of the orbiting asteroid model. MNRAS:508(2):2079-2089

Weber C, Lammer H, Shaikhislamov IF et al (2017) How expanded ionospheres of Hot Jupiters can prevent escape of radio emission generated by the cyclotron maser instability. MNRAS 469:3505–3517

Willes AJ, Wu K (2004) Electron-cyclotron maser emission from white dwarf pairs and white dwarf planetary systems. MNRAS 348:285–296

Willes AJ, Wu K (2005) Radio emissions from terrestrial planets around white dwarfs. Astron Astrophys 432:1091–1100

Wu CS, Lee LC (1979) A theory of the terrestrial kilometric radiation. Astrophys J 230:621–626

Yadav RK, Thorngren DP (2017) Estimating the Magnetic Field Strength in Hot Jupiters. Astrophys J Lett 849:L12 (4pp)

Zarka P, Genova F (1983) Low frequency jovian emission and solar wind magnetic sector structure. Nature 306:767-768

Zarka P (1998) Auroral radio emissions at the outer planets: observations and theories. J Geophys Res 103:20159–20194

Zarka P (2006) Hot jupiters and magnetized stars: giant analogs of the satellite-jupiter system? In: Rucker HO, Kurth WS, Mann G (eds) Planetary radio emissions VI. Austrian Academy of Science Press, Vienna, pp 543–569

Zarka P (2007) Plasma interactions of exoplanets with their parent star and associated radio emissions. Planet Space Sci 55:598–617

Zarka P (2010) Radioastronomy and the study of exoplanets. In: Coudé du Foresto V, Gelino DM, Ribas I (eds) Pathways towards habitable planets, ASP conference series, vol 430. Astronomical Society of the Pacific, San Francisco, pp 175–180

Zarka P, Treumann RA, Ryabov BP, Ryabov VB (2001) Magnetically-driven planetary radio emissions and applications to extrasolar planets. Astrophys Space Sci 277:293–300

Zarka P, Lazio TJW, Hallinan G (2015) Magnetospheric radio emissions from exoplanets with the SKA. In: Advancing astrophysics with the square kilometre array, Giardini Naxos. SKA Organisation (Dolman Scott Ltd), Jodrell Bank Observatory, Macclesfield

Zarka P, Marques MS, Louis C et al (2018) Jupiter radio emission induced by Ganymede and consequences for the radio detection of exoplanets. Astron Astrophys 618:A84

Zarka P, Denis L, Tagger M et al (2020) The low-frequency radio telescope NenuFAR. URSI GASS, Rome, Italy, 29 August - 5 September 2020, https://tinyurl.com/ycocd5ly

Zarka P, Louis CK, Zhang J et al (2024)Location and energy of electrons producing the radio bursts from AD Leo observed by FAST. Astron Astrophys submitted

Zhang J, Tian H, Zarka P et al (2023) Fine structures of radio bursts from flare star AD Leo with FAST observations. Astrophys J 953:65

Zic A, Stewart A, Lenc E et al. (2019) ASKAP detection of periodic and elliptically polarized radio pulses from UV Ceti. MNRAS 488:559–571